\documentclass[conference]{IEEEtran}
\IEEEoverridecommandlockouts
\usepackage{cite}
\usepackage{amsmath,amssymb,amsfonts}
\usepackage{algorithmic}
\usepackage{graphicx}
\usepackage{textcomp}
\usepackage{comment}
\usepackage{xcolor}
\usepackage{subcaption}
\usepackage{multirow}
\usepackage{multicol}
\usepackage{amssymb}
\newcommand{\ceil}[1]{\left\lceil #1 \right\rceil}
\usepackage{lipsum} 
\usepackage{mdframed} 
\usepackage{tabularx}
\usepackage{pifont}
\usepackage{flushend}
\newcolumntype{?}{!{\vrule width 1.5pt}}
\newcommand{\example}[1]{%
 \par\vspace{1pt}%
  \noindent\textbf{Example:} \textit{#1}%
   \par\vspace{1pt}%
}
\usepackage{titlesec}
\usepackage[hidelinks]{hyperref}

\titleformat{\subsubsection}[runin]
  {\normalfont\normalsize\itshape}{\thesubsubsection)}{0.5em}{}[\,\,]

\renewcommand{\thesubsubsection}{\arabic{subsubsection}}

\usepackage{siunitx}
\def\BibTeX{{\rm B\kern-.05em{\sc i\kern-.025em b}\kern-.08em
    T\kern-.1667em\lower.7ex\hbox{E}\kern-.125emX}}
\begin{document}

\title{
A Predictive Approach for Selecting the Best Quantum Solver for an Optimization Problem
\vspace{-10pt}}
\author{\IEEEauthorblockN{Deborah Volpe\IEEEauthorrefmark{1},
Nils Quetschlich\IEEEauthorrefmark{2}, Mariagrazia Graziano\IEEEauthorrefmark{3},  Giovanna Turvani\IEEEauthorrefmark{1}, and Robert Wille\IEEEauthorrefmark{2}\IEEEauthorrefmark{4} }
\IEEEauthorblockA{\IEEEauthorrefmark{1}Department of Electronics and Telecommunications,
Politecnico di Torino
Italy\\
\IEEEauthorrefmark{2}Chair for Design Automation, Technical University of Munich, Germany\\
\IEEEauthorrefmark{3}Department of Applied Science and Technology,
Politecnico di Torino
Italy\\
\IEEEauthorrefmark{4}Software Competence Center Hagenberg GmbH (SCCH), Austria\\
 \href{mailto:deborah.volpe@polito.it}{deborah.volpe@polito.it},
\href{mailto:nils.quetschlich@tum.de}{nils.quetschlich@tum.de},
\href{mailto:mariagrazia.graziano@polito.it}{mariagrazia.graziano@polito.it},\\
\href{mailto:giovanna.turvani@polito.it}{giovanna.turvani@polito.it} and \href{mailto:robert.wille@tum.de}{robert.wille@tum.de}} \vspace{-32pt}} 
\maketitle

\begin{abstract}
Leveraging quantum computers for optimization problems holds promise across various application domains. Nevertheless, utilizing respective quantum computing solvers requires describing the optimization problem according to the \textit{Quadratic Unconstrained Binary Optimization} (QUBO) formalism and selecting a proper solver for the application of interest with a reasonable setting. Both demand significant proficiency in quantum computing, QUBO formulation, and quantum solvers, a background that usually cannot be assumed by end users who are domain experts rather than quantum computing specialists.  
While tools aid in QUBO formulations, support for selecting the \mbox{best-solving} approach remains absent.
 This becomes even more challenging because selecting the best solver for a problem heavily depends on the problem itself.

In this work, we are accepting this challenge and propose a predictive selection approach, which aids end users in this task. To this end, the solver selection task is first formulated as a classification task that is suitable to be solved by supervised machine learning. Based on that, we then propose strategies for adjusting solver parameters based on problem size and characteristics.

Experimental evaluations, considering more than 500 different QUBO problems, confirm the benefits of the proposed solution. In fact, we show that in more than 70\% of the cases, the best solver is selected, and in about 90\% of the problems, a solver in the top two, i.e., the best or its closest suboptimum,  is selected.

This exploration proves the potential of machine learning in quantum solver selection and lays the foundations for its automation, broadening access to quantum optimization for a wider range of users. 

The pre-trained classifier is integrated into the MQT Quantum Auto Optimizer (MQT QAO) framework,  publicly available on GitHub (\url{https://github.com/cda-tum/mqt-qao}) as part of the Munich Quantum Toolkit (MQT).
\end{abstract}

\vspace{-12pt}
\section{Introduction}
\vspace{-8pt}
\textit{Quantum computing} holds promise for solving optimization problems across various application fields, including finance~\cite{lang2022strategic},  resource allocation~\cite{ohyama2023resource},  and scheduling~\cite{zhang2022solving}. However, harnessing \textit{quantum optimization}---i.e., solving optimization problems leveraging quantum computers---demands \textit{expertise} in quantum computing, especially when formulating optimization problems in a format suitable for quantum computers such as QUBO and selecting the most promising solver to determine the result. 

This poses a challenge for researchers and industries not directly involved in quantum computation, hindering their initial exploration of quantum solutions for \mbox{real-world} applications.

While some frameworks for supporting the translation of optimization problems into a quantum-compliant format exist, such as that proposed in~\cite{autoqubo-repository, moraglio2022autoqubo, pauckert2023autoqubo,xavier2023qubo, volpe2024towards, quetschlich2023towards}, the literature lacks tools to assist in leveraging \textit{quantum solvers}, including both selecting the \textit{most promising solver} and its respective \textit{parameters}. Unfortunately, executing and evaluating all quantum solvers and their parameters configurations can be expensive both in terms of time and money, especially considering the limitations of nowaday quantum computers, making this naive approach unfeasible. For these reasons, \textit{automating} solver selection and its parameter configuration can significantly streamline the process, enabling users to achieve good results with minimal effort and expanding access to quantum optimization for a broader range of users. 

This work proposes approaching the challenge of selecting solvers as a \textit{classification} task, considering the most popular quantum solvers---i.e., \textit{Quantum Annealer},  \textit{Quantum Approximate Optimization Algorithm}, \textit{Variational Quantum Eigensolver}, and \textit{Grover Adaptive Search}---and \textit{supervised machine learning models}.  In addition to quantum solvers, the model also considers the classical \textit{Simulated Annealing}, the classical counterpart of Quantum Annealing. By including a classical solver, we demonstrate the versatility of the approach and that the model could be further expanded by adding more classical and quantum solvers. In this way, the classifier can identify the optimal quantum solver and distinguish problems that could gain advantages from quantum exploration compared to classical methods~\cite{chancellor2017modernizing}.
Moreover, in this article, some techniques for \textit{adjusting solver parameters} based on problem size
and characteristics are proposed. 

The considered dataset has been obtained by solving more than \textit{500} different QUBO problems. The obtained results are promising since the solvers' setting guarantees, on average, promising results, and the best solver is identified in \textit{more than 70\%} of the cases. Moreover, a solver in \textit{the top two}, i.e., the best or its closest suboptimum,  is selected in \textit{about 90\%} of the considered problems.

This exploration proves the potential of supervised machine learning in selecting the best solver for a problem, significantly enhancing the accessibility of quantum optimization across a wider range of users. 
The predictor streamlines the solver selection process, hiding its complexity into a black box solution. Consequently, end users can effortlessly evaluate the quantum solution's effectiveness for their application without deep quantum solver expertise. Moreover, this approach mitigates the resource-intensive task of executing multiple solvers in terms of time and cost, providing a more efficient pathway to identify potential quantum advantages. The pre-trained classifier is integrated into the MQT Quantum Auto Optimizer (MQT QAO) framework,  publicly available on GitHub (\url{https://github.com/cda-tum/mqt-qao}) as part of the Munich Quantum Toolkit (MQT, \cite{willeMQTHandbookSummary2024}).

The rest of the article is organized as follows. 
Section~\ref{sec:SolvingOptimizationProblems} briefly reviews quantum optimization, focusing on the workflow needed for solving an optimization problem with quantum computers and quantum solvers. Section~\ref{sec:MotivationAndIdea} outlines the motivation behind this work and discusses the general idea. The proposed supervised learning approach for solver selection and the considered solver setting are presented in  Section~\ref{sec:SupervisedLearningApproach}, focusing on the methodology adopted for the classification model development. The obtained results are provided in Section~\ref{sec:Results}, while the effectiveness of the proposed approach is examined in Section~\ref{sec:Discussion}. Finally, in Section~\ref{sec:conclusions}, conclusions are drawn, and future perspectives are illustrated.

\section{Solving Optimization Problems\\ with Quantum Computers}\label{sec:SolvingOptimizationProblems}
This section reviews quantum optimization, concentrating on the \mbox{quantum-compliant} problem formulation and quantum solvers. Moreover, the main steps required for solving generic optimization problems with quantum computers are summarized.

To this end, two different paradigms of quantum computing exploited to solve optimization problems are considered: 
\begin{itemize}
    \item quantum annealing, implemented through a \textit{\mbox{special-purpose} quantum computer} called \textit{quantum annealer}~\cite{kadowaki1998quantum,chakrabarti2023quantum,rajak2023quantum, venturelli2015quantum, ManufacturedSpins};
    \item \textit{quantum circuit model}~\cite{nielsen_quantum_2010}, defining proper algorithms involving both classical and quantum or only \textit{quantum  computers \mbox{general-purpose}}.
\end{itemize}
However, independently from the exploited paradigm, the problem must be written into a compliant formulation involving only binary variables.

\subsection{QUBO Model}
The usual formulation for solving an optimization problem leveraging quantum computers is the \textit{Quadratic Unconstrained Binary Optimization} (QUBO) model~\cite{glover2018tutorial,combarro2023practical}. This formulation describes the problem involving only binary variables. The \textit{quadratic} term refers to the highest power applied to these binary variables, and \textit{unconstrained} indicates that the constraints cannot be considered conventionally, but only via introduction of \textit{quadratic penalties to the objective function}. Therefore, the corresponding objective function can be written as
\begin{equation}
	f(\textbf{x}) = c+ \sum_{i} x_{i} \cdot a_{i} + \sum_{i<j} b_{ij} \cdot x_{i}x_{j} \, ,
\end{equation}
where $x_{i} \in [0,1]$ is a binary variable, $x_{i}x_{j}$ is a coupler of two variables, $a_{i}$ is a single-variable weight, $b_{ij}$ is a strength that controls the influence of variables $i$ and $j$, and $c$ is an offset, which can be neglected during the optimization since it shifts the entire objective function without altering the extremal points.



\subsection{Quantum Solvers} \label{sec:Quantum solver}
Different quantum solvers have been proposed for optimization problems provided according to the QUBO model. The most popular are briefly discussed in the following. 
\subsubsection{Quantum Annealer} 
\begin{figure}[h]
    \centering\vspace{-10pt}
    \includegraphics[width=\linewidth]{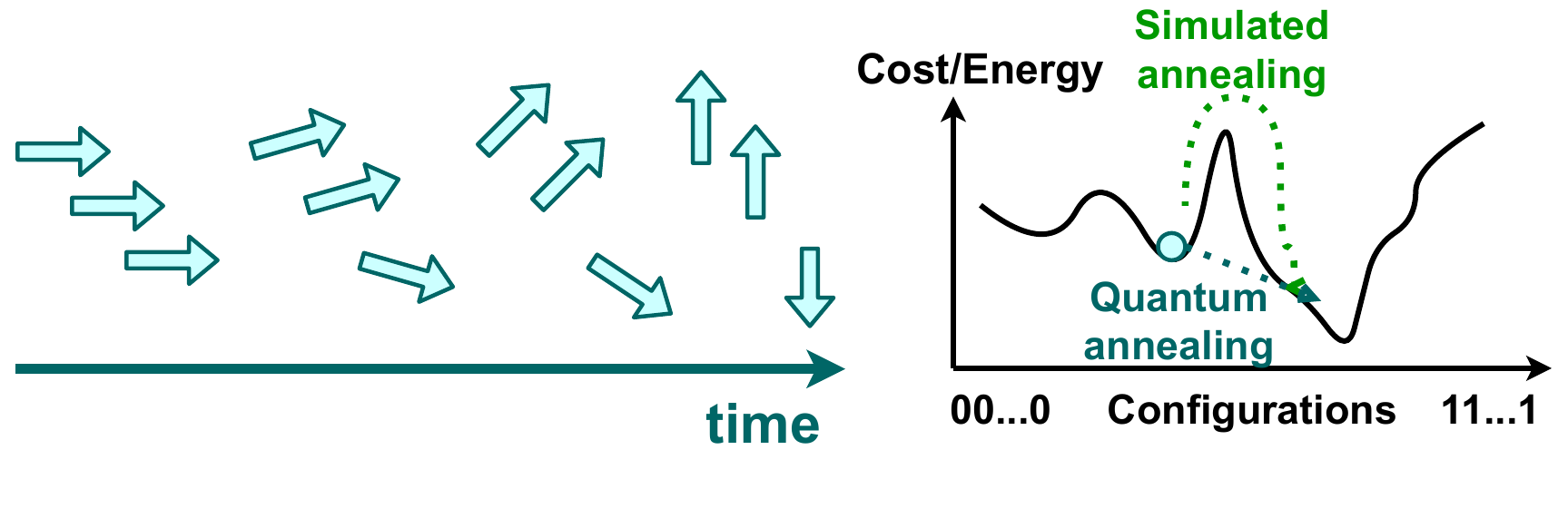}\vspace{-15pt}
    \caption{Evolution of the Quantum Annealer system and comparison against Simulated Annealing exploration mechanism.\vspace{-8pt}}
    \label{fig:QuantumAnnealer}
\end{figure}

(QA, ~\cite{kadowaki1998quantum, venturelli2015quantum, ManufacturedSpins}) is a \mbox{special-purpose} quantum computer designed to solve optimization problems with the quantum annealing algorithm by harnessing \textit{quantum superposition} and \textit{tunneling} to explore the solution space of a given optimization problem.\\
It implements the quantum counterpart of a \textit{Simulated Annealing} (SA, ~\cite{Kirkpatrick1983}) algorithm, which solves optimization problems by encoding them in the energy of a system whose thermal evolution for reaching the ground state is emulated to obtain the optimal solution.
Analogously, in QA, the quantum system begins in a superposition of states and evolves following the principles of the quantum adiabatic theorem towards its lowest energy state, i.e., the problem's optimal solution. The system evolution, shown in Figure~\ref{fig:QuantumAnnealer}, is managed by a schedule parameter known as the \textit{annealing time} or \textit{annealing schedule}, which determines the rate at which the system transitions from the initial to the final state. Properly setting this parameter is crucial to ensure \textit{adiabaticity}, which entails maintaining the system's ground state throughout the evolution process and achieving high-quality solutions.
\\Quantum annealers comprise a set of qubits interconnected via tunable couplers, forming a network known as the \textit{annealing graph}. The connectivity of this graph is constrained by the qubit technology. A \textit{minor embedding}~\cite{konz2021embedding} step is employed to address this limitation and allows the mapping of a fully-connected problem onto a partially-connected device. It maps each QUBO variable to a set of strictly correlated physical qubits, overcoming the connectivity constraint. This implies that the number of qubits required for solving an optimization problem could be significantly higher than the binary variables involved.
\vspace{-3pt}
\subsubsection{Quantum Approximate Optimization Algorithm} 
\begin{figure}[h]
    \centering\vspace{-7pt}
    \includegraphics[width=1\linewidth]{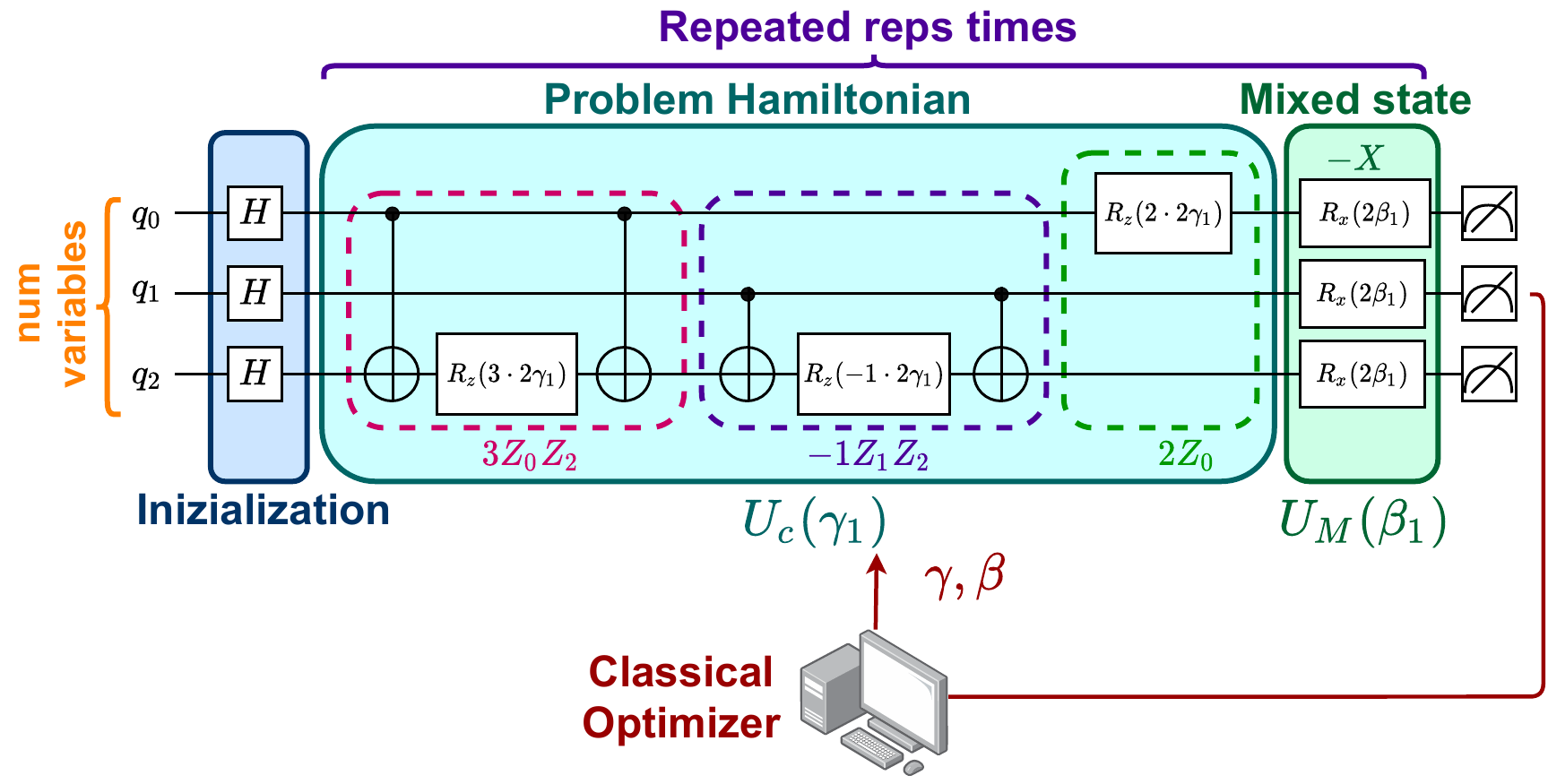}\vspace{-4pt}
    \caption{Quantum Approximate Optimization Algorithm.  \vspace{-7pt}}
    \label{fig:QAOAExample}
\end{figure}

(QAOA, ~\cite{blekos2023review,farhi2014quantum}) is a hybrid \mbox{quantum-classical} technique for solving optimization problems.  It is a \textit{variational algorithm}, i.e., an iterative procedure applied to some parameters of the quantum circuit, such as the rotation amounts of single-qubit gates. These parameters are optimized classically such that the final quantum state encodes the optimal solution of the problem. The variational circuit provides an overall unitary evolution composed of a layer depending on problem Hamiltonian and a mixed state Hamiltonian, as shown in Figure~\ref{fig:QAOAExample}. This is chosen to mimic quantum adiabatic evolution behavior through Trotterization, a technique for approximating the continued evolution with discretized steps. This is done by decomposing the evolution operator into a series of simpler time-independent operators by using the Trotter-Suzuki method \cite{suzuki1976generalized}.  An initialization state must be obtained with a state preparation circuit to guarantee a proper evolution. Therefore, the key parameters of the algorithm are the number of steps for Trotter discretization, called repetitions (reps), the state preparation circuit, the mixed state, and the classical optimizer. This solver demands the execution of circuits with a number of qubits equal to the number of QUBO variables.

\subsubsection{Variational Quantum Eigensolver}  
\begin{figure}[h]
    \centering \vspace{-7pt}
    \includegraphics[width=1\linewidth]{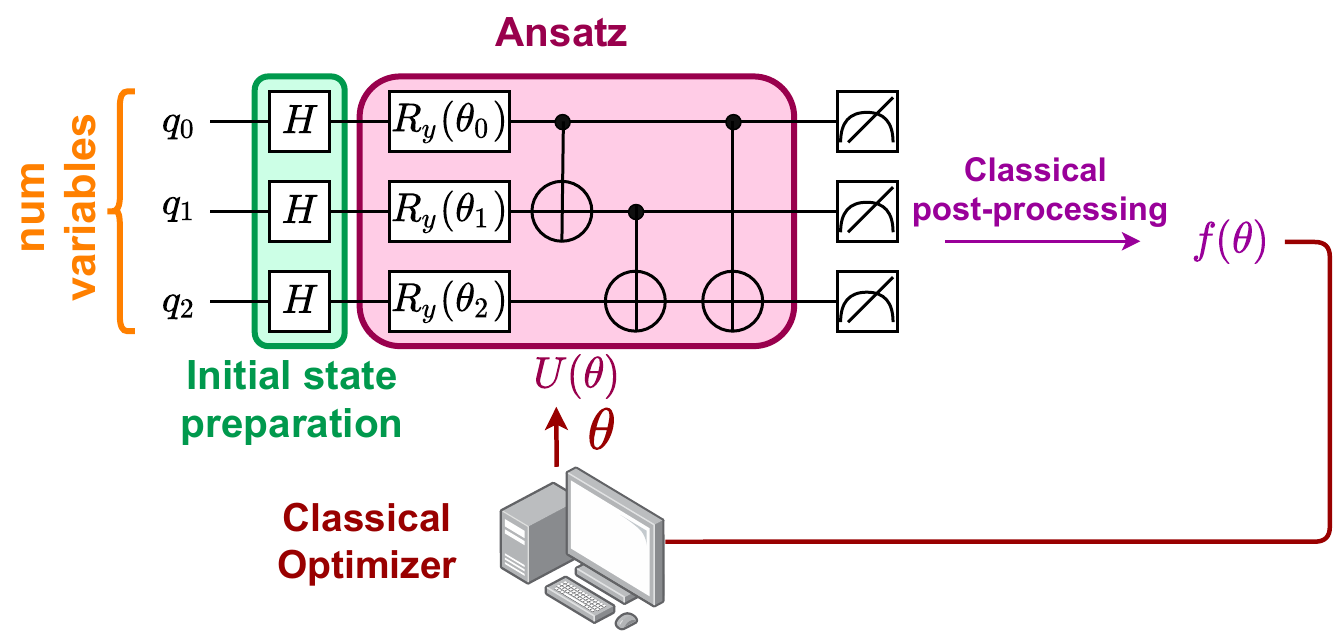} \vspace{-4pt}
    \caption{Variational Quantum Eigensolver. \vspace{-7pt}}
    \label{fig:VQEExample}
\end{figure}

(VQE, ~\cite{tilly2022variational, peruzzo2014variational}) is a hybrid \mbox{quantum-classical} algorithm that aims to minimize a scalar cost function mapped onto a Hamiltonian. It is a variational algorithm in which the quantum circuit, as shown in Figure~\ref{fig:VQEExample}, is composed of an ansatz whose parameters are properly optimized by a classical solver such that the expectation value of the system with respect to the Hamiltonian of the problem is minimized.  Therefore, the key parameters of the algorithms are the ansatz and the classical optimizer. This solver requires the execution of circuits with a number of qubits equal to the number of QUBO variables.
\vspace{10pt}
\subsubsection{Grover Adaptive Search}  
\begin{figure}[h]
    \centering \vspace{-12pt}
    \includegraphics[width=\linewidth]{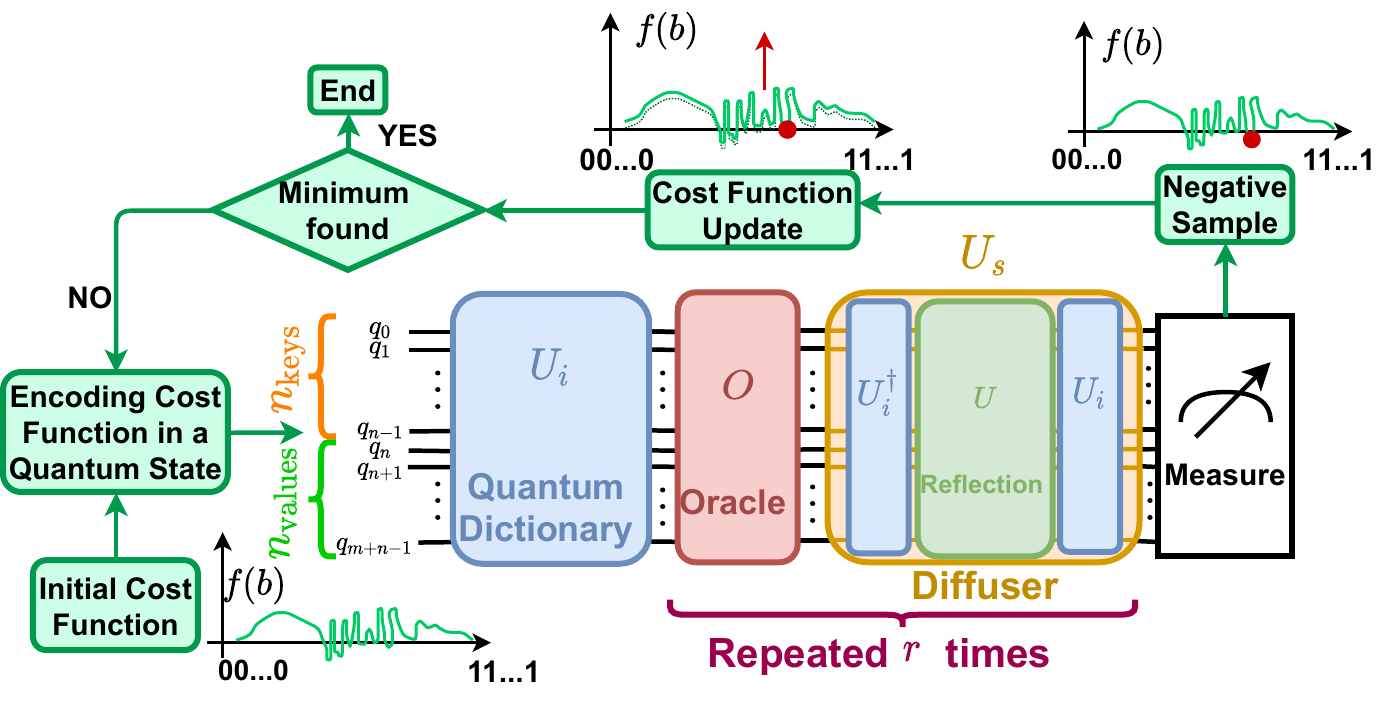} \vspace{-20pt}
    \caption{Grover Adaptive Search \vspace{-20pt}}
    \label{fig:GASScheme}
\end{figure}

(GAS,~\cite{bulger2003implementing, gilliam2021grover, sano2023accelerating, sano2023qubit, giuffrida2022engineering}) is a hybrid \mbox{quantum-classical} algorithm that leverages a successive approximation approach to minimize a cost function. In particular,  the algorithm works by repeatedly sampling negative values of the objective function and then incrementally shifting it upward by the same amount until no further negative values are observed. Therefore, the last negative sample represents the minimum of the objective function. Negative samples are obtained through the \textit{Grover Search} routine as shown in Figure~\ref{fig:GASScheme}, where the cost function is encoded onto the quantum state using a quantum dictionary, which allows the expression of a key (function domain) and value (function image) relation. Consequently, the exploited quantum circuit requires a number of qubits equal to $n_{\textrm{keys}} + n_{\textrm{values}} $, where $n_{\textrm{keys}}$  is the number of binary variables and $n_{\textrm{values}}$ is the number of bits necessary for representing the image of the function. Selecting the appropriate $n_{\textrm{values}}$ poses an initial challenge in utilizing this algorithm since it necessitates knowledge of the function's operational range. \\
Since the optimal number of Grover rotations $r$ depends on the number of negative samples over the entire solution space that cannot be predetermined, techniques have been proposed in~\cite{baritompa2005grover,gilliam2021grover,giuffrida2022engineering} for choosing it in each algorithm step. Another critical aspect of the algorithm involves determining whether negative values remain. Although obtaining a positive sample via Grover search is an event that occurs when \mbox{non-negative} function values are available, this cannot serve as a direct terminating condition, as it may also occur when the number of Grover rotations is improperly set. Consequently, the standard approach is to count the number of consecutive measured positive samples and stop the algorithm when this number overcomes a certain threshold $th$ to be properly chosen.

\subsection{From Optimization Problems to Quantum Solutions}
\begin{figure*}[t]
	\centering
	\includegraphics[width=1\textwidth]{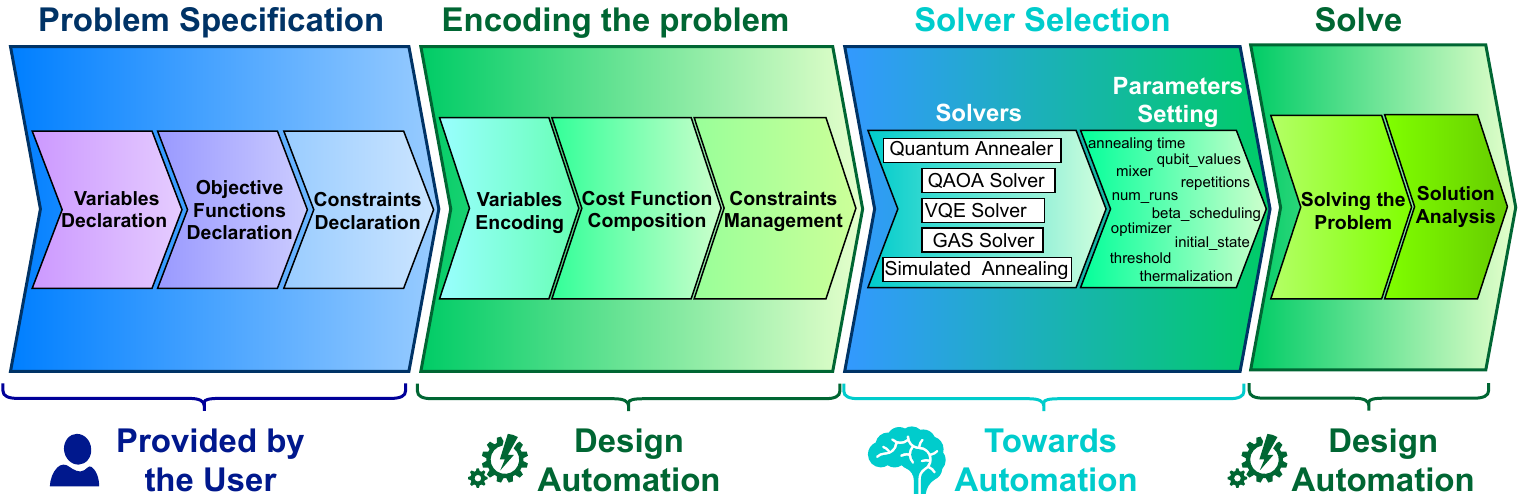} \vspace{-20pt}
\caption{Quantum optimization flow. \vspace{-20pt}}
\label{fig:ASS}
\end{figure*}

Using the basis and methods revised above, the workflow shown in Figure~\ref{fig:ASS} emerged, enabling end users to solve optimization problems leveraging quantum computers or quantum annealers. More precisely the steps needed for harnessing quantum solvers are discussed in the following. 

\subsubsection{Problem specification:}
The initial step of the optimization process involves defining the problem specifications, which include the \textit{variables}, \textit{objective functions} (criteria for optimization), and any \textit{constraints} required for a valid solution. \\
These variables can take on \textit{binary}, \textit{discrete}, or \textit{continuous} values, delineating their characteristics such as unipolar/bipolar nature or defined value ranges. The objective function represents a parametric description of a figure of merit, with its optimal value either being the minimum or maximum, necessitating the specification of the optimization direction. 

\subsubsection{Encoding the Problem: } \label{sec:writing}
Afterwards, the problem has to be described in a \textit{solver-compliant} form, particularly employing the QUBO model. As mentioned, this model exclusively supports \textit{binary variables}, requiring proper \textit{encoding mechanisms} for expressing continuous and discrete ones. In the case of \mbox{multi-objective} optimization, the aggregation approach can be employed to merge objective functions into a higher scalar function, which reflects a preference criterion, \textit{composing the cost function}. Moreover, \textit{constraints} can be managed only as \textit{weighted penalty functions} incorporated within the primary objective function. Finally, any eventual \mbox{higher-order} polynomials need to be reduced to second-order terms.

\subsubsection{Solver Selection:}
Following the problem specification stage, the next step involves selecting a solver, such as QA, QAOA, VQE, or GAS. Each of these optimizers has a peculiar exploration mechanism. Therefore, different performance is expected when considering a solver instead of another.  
Subsequently, careful attention must be paid to selecting suitable settings for the chosen solver. For example, in the case of GAS, a proper number of qubits for the value and the threshold has to be selected.

\subsubsection{Solve: }\label{sec:execute}
The resulting QUBO formulation is then submitted to the solver with an \textit{appropriate parameter configuration} (Solving the Problem). Due to their stochastic nature, solvers are typically executed multiple times, and the best result is considered. Access to quantum devices is possible via \textit{cloud services}, through an account. Otherwise, quantum solvers such as QAOA, VQE, and GAS can also be executed by exploiting \textit{classical simulators}, as those presented in~\cite{fingerhuth2018open, guerreschi2020intel, zulehner2018advanced, hillmich2020just, vincent2022jet, villalonga2019flexible}.\\
Finally, the acquired solution must be decoded to retrieve the original problem variables; then, its quality must be assessed (Solution analysis). This entails evaluating the initial cost functions using the obtained configuration and verifying constraint satisfaction.
\vspace{-8pt}

\subsection{Existing Tools}

These steps are rather complex and can easily overwhelm users who are experts in their respective domains (such as finance~\cite{lang2022strategic},  resource allocation~\cite{ohyama2023resource},  and scheduling~\cite{zhang2022solving}) rather than in quantum computing. Therefore, this procedure may benefit from design automation to enhance the accessibility of quantum solutions. 

In recent years, several libraries and some tools have emerged to streamline the QUBO formulation. The main libraries include \textit{pyqubo}~\cite{pyqubo-docs, zaman2021pyqubo}, \textit{qubovert}~\cite{qubovert-docs}, \textit{dimod}~\cite{dimod-docs}, \textit{Qiskit-optimization}~\cite{qiskit-optimization-docs}, \textit{Fixstarts Amplify}~\cite{fixstarts-docs} and \textit{openQAOA Entropica}~\cite{openQAOA-entropica-labs-docs}.
They significantly facilitate complex aspects of the QUBO formulation process. Nonetheless, their primary limitation lies in their lack of support for fully automating these steps, thereby restricting their accessibility to users with at least a minimum level of expertise in the field. 
Moreover,  in response to user demands for comprehensive automation, three automation frameworks have surfaced in the last two years: \textit{AutoQUBO}~\cite{autoqubo-repository, moraglio2022autoqubo, pauckert2023autoqubo}, QUBO.jl~\cite{xavier2023qubo}, and MQT QAO \cite{volpe2024towards}. Their focus primarily lies in assisting with the problem encoding stage and interface with solvers. Finally, tools and methods exist that guide end-users through the corresponding compilation process~\cite{quetschlich2024mqtpredictor}.

Nevertheless, to our knowledge, no tool is currently available in the literature to aid in the solver selection stage despite its not negligible complexity, especially for non-expert users, since it requires deep knowledge of quantum solvers, which may inhibit their interest in experimenting with quantum solutions. Addressing this gap remains an ongoing challenge that we aim to address in this manuscript. 

\begin{figure*}[t]
\begin{subfigure}[t]{0.66\columnwidth}
	    \centering
\includegraphics[width=\textwidth]{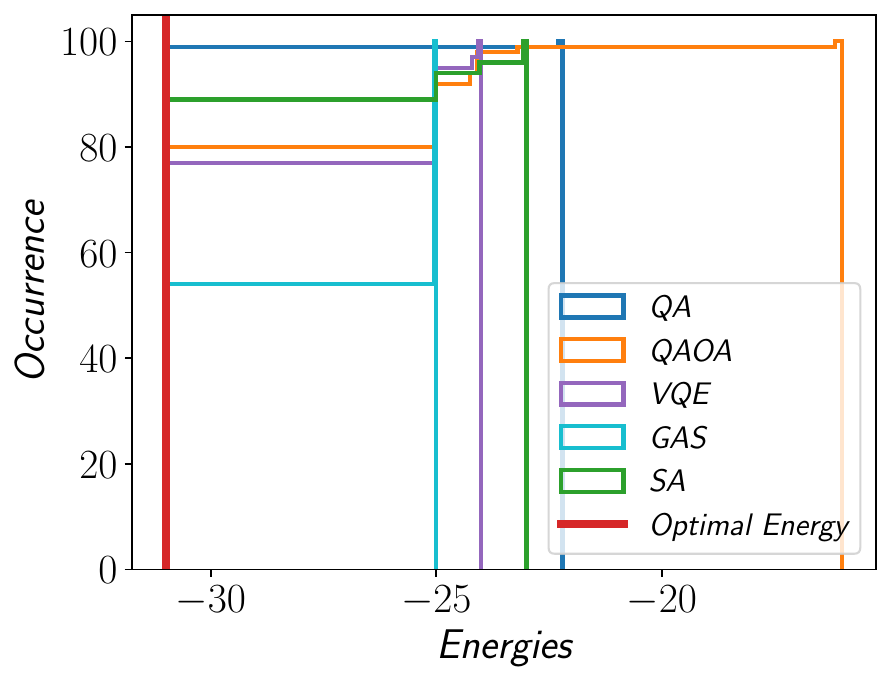}
	    \caption{ Cumulative distribution of a Set Packing problem, where QA emerges as better solver  }
	    \label{fig:QAExample}
	\end{subfigure}
 \quad
	\begin{subfigure}[t]{0.66\columnwidth}
	    \centering
	    \includegraphics[width=\textwidth]{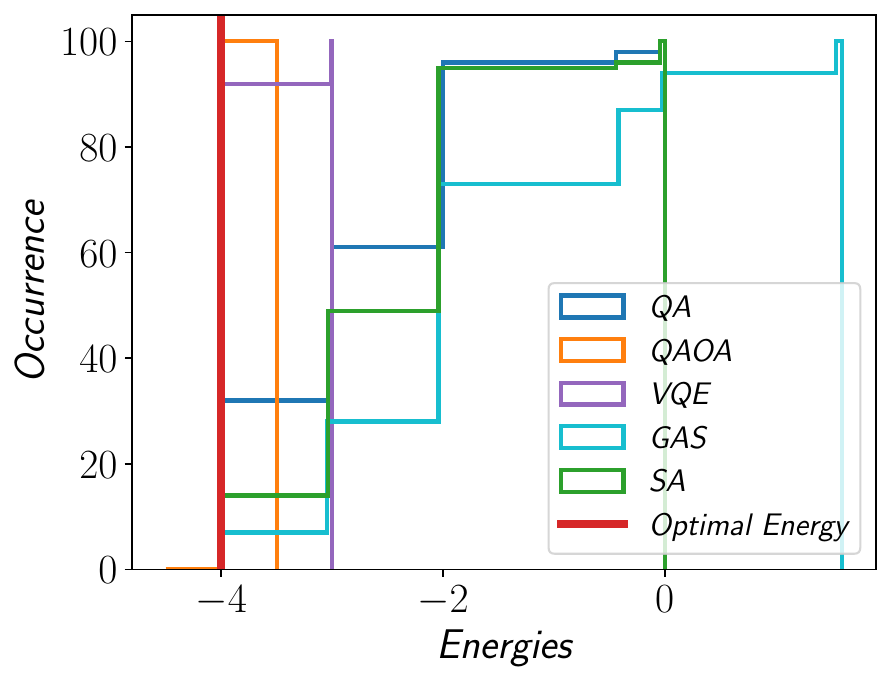}
	    \caption{Cumulative distribution of a Kclique problem, where QAOA emerges as better solver }
	    \label{fig:QAOAExampleCumulative}
	\end{subfigure}
 \quad
        \begin{subfigure}[t]{0.66\columnwidth}
	    \centering
	    \includegraphics[width=\textwidth]{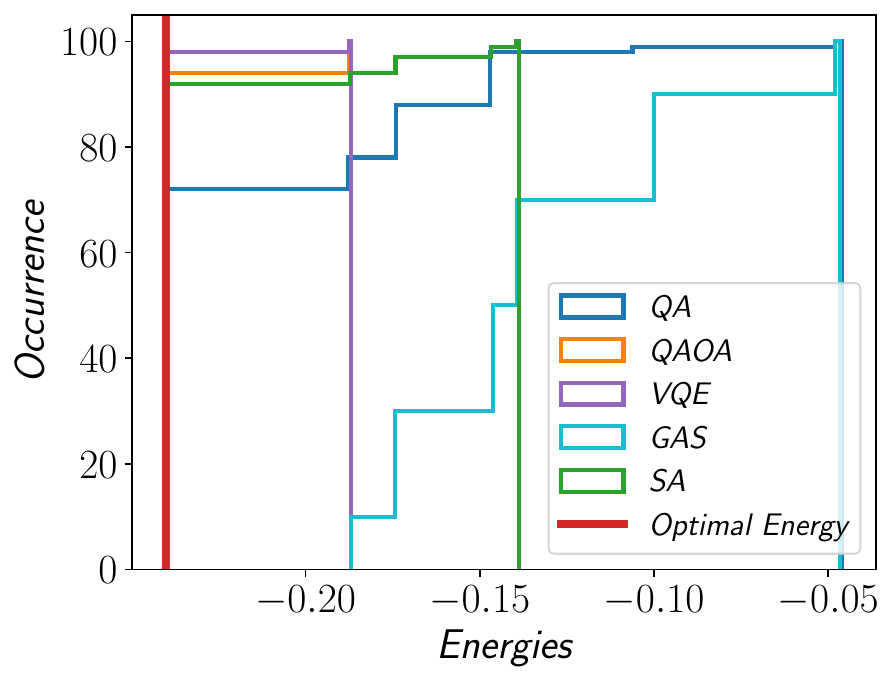}
	    \caption{Cumulative distribution of a Portfolio Optimization problem, where VQE emerges as better solver }
	    \label{fig:VQEExampleCumulative}
	\end{subfigure}
 \begin{center}
 	\begin{subfigure}[t]{0.66\columnwidth}
	    \centering
	    \includegraphics[width=\textwidth]{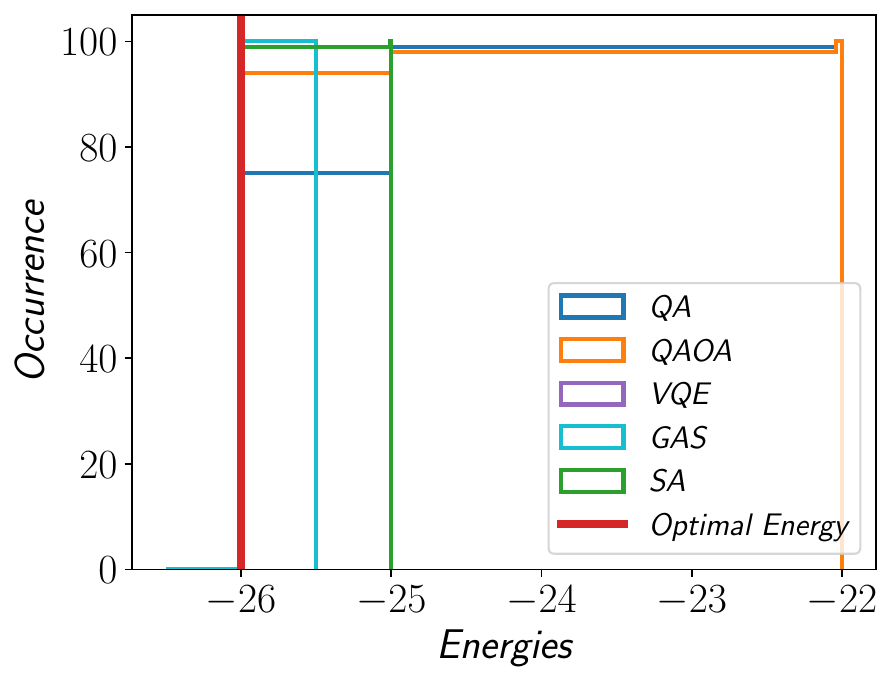}
	    \caption{Cumulative distribution of a Maxcut problem, where GAS emerges as better solver}
	    \label{fig:GASExampleCumulative}
	\end{subfigure}
 \quad\quad
        \begin{subfigure}[t]{0.66\columnwidth}
	    \centering
	    \includegraphics[width=\textwidth]{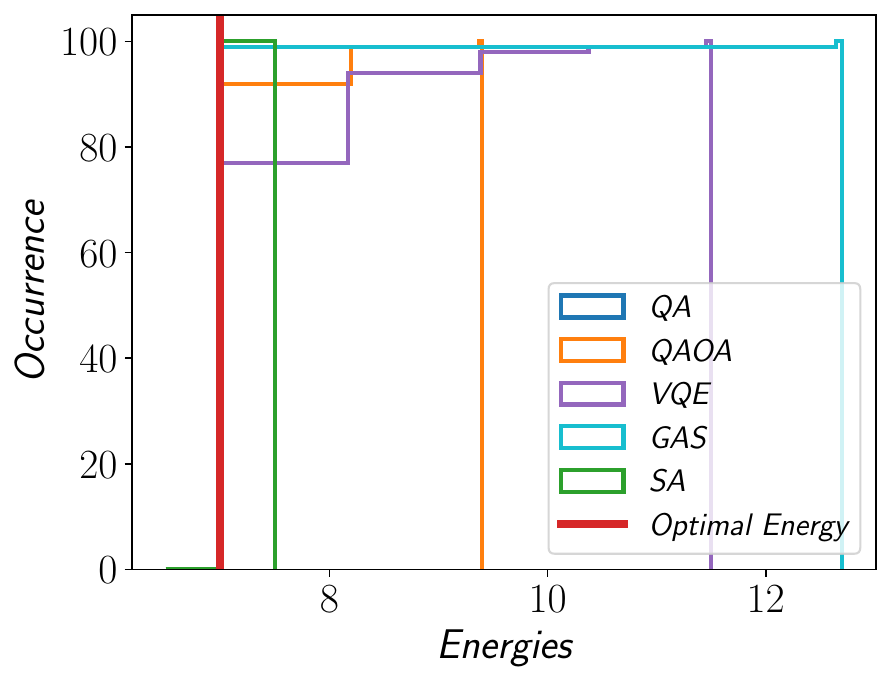}
	    \caption{Cumulative distribution of a Minimum Vertex Cover problem, where SA emerges as better solver}
	    \label{fig:SAExampleCumulative}
	\end{subfigure}
  \end{center}
	\caption{These figures show cumulative distributions obtained by solving different QUBO problems with QA, QAOA, VQE, GAS, and SA. To understand the meaning of these plots, one rule has to be considered: the probability of obtaining the optimal value (or a value close to it) with a specific solver is higher when its corresponding cumulative distribution is more concentrated on the left of the plot, where the lowest values of the objective function are located. It is possible to notice that the best solver, in terms of solution quality, varies across each problem examined. This observation highlights the dependence of solver effectiveness on the problem itself. To underline this argument, we have evaluated applications from various domains, leading to completely different results in terms of optimal solver for each particular case.
\vspace{-15pt}
 }
	\label{fig:UseCase}
\end{figure*}

\vspace{-5pt}
\section{Towards Automation of\\ Quantum Solvers Selection } \label{sec:MotivationAndIdea}
\vspace{-3pt}
This section discussed the problem selection challenges, including both the choice of the optimization solver and the setting of its parameters, for end-users and the unmet needs that this work aims to address.
\vspace{-5pt}
\begin{figure*}[t]
	\centering
	\includegraphics[width=1\textwidth]{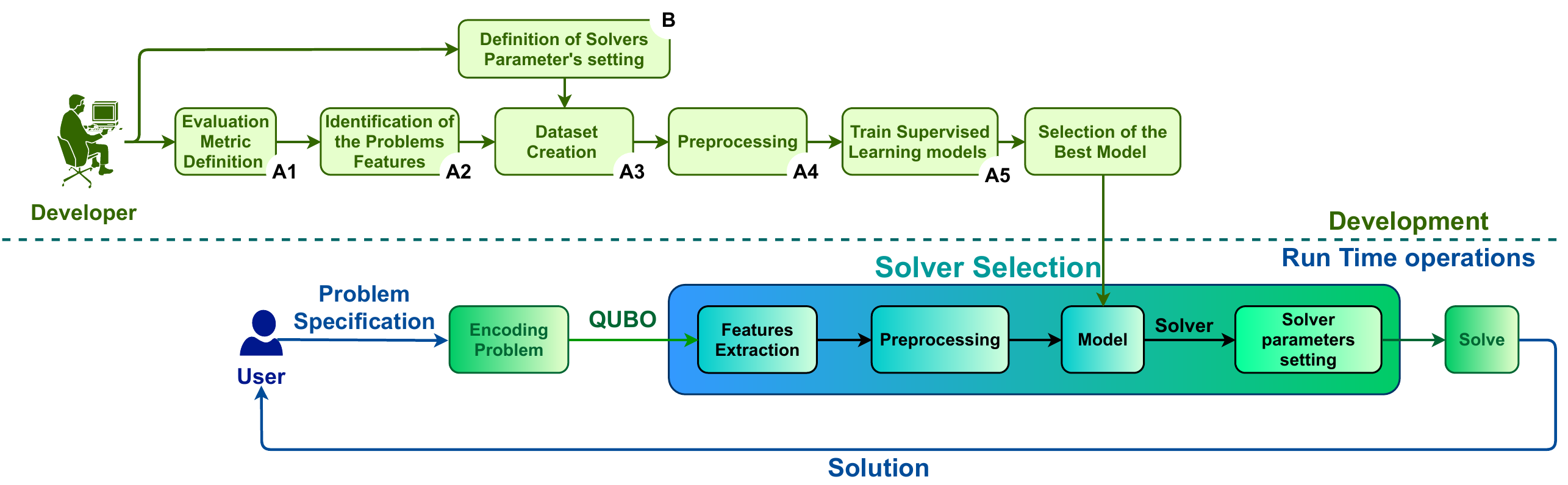} \vspace{-10pt}
\caption{Workflow for implementation of the proposed predictive model and for its exploitation from the user perspective. The letters and numbers correspond to the subsections where each step is discussed. \vspace{-15pt}}
\label{fig:Flow}
\end{figure*}
\vspace{-3pt}
\subsection{Considered Challenge}
\vspace{-3pt}

 Estimating a priori what might be the \textit{best solver for the problem of interest} or whether a classical or quantum approach might be better is a non-trivial task.
 
 This problem is similar to the \textit{No Free Lunch Theorem}~\cite{adam2019no}, which declares that every optimization technique performs as well as every other one on average, and \textit{a single best optimization algorithm does not exist}. Consequently, \textit{different optimization problems correspond to different optimal solvers}, complicating a priori selection of the most suitable solver for a specific problem.

 \vspace{2pt}
 \example{
 The problem is illustrated by the cumulative plots reported in Figure~\ref{fig:UseCase}, considering five different optimization problems as examples. Cumulative plots are a common representation of heuristic solver results, showing the distribution of solver outcomes across multiple runs. To correctly interpret the plot, a simple rule has to be considered: the probability of obtaining the optimal value (or a value close to it) with a specific solver is higher when its corresponding cumulative distribution is more concentrated on the left of the plot, where the lowest values of the objective function are located. These examples show that each solver is the best for one problem, and a single best optimization approach does not exist. }
 
 However, executing and evaluating all quantum solvers can be expensive in terms of time and money, particularly given the current limitations. Nowadays, running quantum circuit model solvers on simulators is necessary to get results not affected by errors introduced by hardware noise and non-ideality phenomena, such as qubit relaxation and decoherence. Moreover, accessing real quantum computers often entails subscription fees that cannot be neglected. These problems could inhibit the exploration of quantum solutions for \mbox{real-world} optimization scenarios. 

  Furthermore, properly \textit{configuring quantum solver} parameters demands significant expertise in quantum computing, which is unusual among conventional optimization users. Also, in this case,  assessing all possible parameters combinations to find the optimal configuration is prohibitively costly and impractical, especially given the current constraints on access to quantum resources and quantum simulators.

  Moreover, adjusting scaling factors such as annealing time in QA or determining parameters like the number of qubits for value representation in GAS could also be challenging even for users well-versed in quantum computing. Indeed, in the former case, it necessitates leveraging prior experience, while estimating the operational range of the function is required in the latter.

For all these reasons, the solver selection step can greatly benefit from \textit{automation} for both choosing the best solver and setting its parameters according to the specific problem to solve. 
\vspace{-3pt}
\subsection{General Idea} \label{sec:GeneralIdea}
\vspace{-3pt}
This work aims to offer tools for assisting non-expert users in \textit{selecting quantum solvers and configuring their parameters} for any QUBO-representable optimization problem, hiding the complexities inherent in these tasks from the users. 

For solver selection, we propose to leverage \textit{supervised machine learning}. 
In order to develop this solution, the following steps are necessary:
\begin{enumerate}
    \item Determining an \textit{evaluation metric} for defining the best solver.
    \item Extracting and identifying meaningful information from the QUBO formulation to serve as \textit{features} characterizing the underlying optimization problem in the classification task.
    \item Creating the \textit{dataset}, which consists of:
    \begin{enumerate}
        \item selecting a diverse set of benchmarks;
        \item defining a strategy for handling parity cases;
        \item analyzing the dataset's significance.
    \end{enumerate}
    \item Applying data \textit{preprocessing} techniques.
    \item Evaluating performance with various \textit{classification models} and selecting the most effective.
    \item Finally, obtaining the trained models and exploiting them to predict the best solver for new problems runtime. 
\end{enumerate}

For configuring solver parameters, we propose leveraging experimental results from the literature to identify scaling trends and \mbox{state-of-the-art} techniques for extracting QUBO information from its formulation. For instance, to determine the number of repetitions in QAOA, we suggest using information/insights gained by plots available in the literature, like those in~\cite{lee2021parameters,niu2019optimizing},  that depict the empirical scaling of this parameter in experimental results. This information can then be used to estimate a reasonable value based on the problem size. These parameters scaling can be further improved by executing additional tests. 

\section{A Supervised Learning Approach \\for Solver Selection}\label{sec:SupervisedLearningApproach}
Running all possible quantum solvers with even different parameters easily exceeds what is practically feasible due to the required effort to implement them and the associated costs for their execution. Therefore, in this work, we propose a methodology for \textit{predicting the best solvers} among the available ones without explicitly executing all of them. In this article, we consider QA, QAOA, VQE, and GAS as quantum solvers and SA as a classical counterpart.  In order to reach the target, the selection of the solver problem is interpreted as a classification task, addressable by exploiting \textit{supervised machine learning techniques}. Furthermore,  simple strategies for effectively configuring and scaling the parameters of quantum solvers are proposed.


From the users' perspective, the provided tool, integrated into the MQT Quantum Auto Optimizer framework that implements the problem translation, enables the achievement of good results without trying out all possible quantum solvers, even with different parameters, and any knowledge about quantum computing. Indeed, it can be seen as a black-box tool (the box in Figure~\ref{fig:Flow}) that automatically makes those decisions---in real-time and without any associated costs of actually executing any circuit and hiding the required steps from the users. The tool can then substitute the former Solver Selection step with the two tasks, as shown in Figure~\ref{fig:ASS}. However, realizing this black box requires significant effort on the developer's side. What this exactly means is explained in the following and shown in Figure~\ref{fig:Flow}.

\subsection{Prediction of the Best Solver}
As anticipated in Section~\ref{sec:GeneralIdea}, several steps are required for developing a solver predictor. They are discussed in the following.  
\subsubsection{Evaluation Metric:}
First of all, establishing an \textit{evaluation metric}, i.e., a singular score determining the best solver for each problem, is essential for developing the solver predictor.  In general, an optimization problem solver is evaluated based on the \textit{quality of the obtained solution} and the \textit{time required to achieve convergence}. However, comparing convergence times among solvers in this context is non-trivial.  Indeed, the currently available quantum circuit model devices do not allow reliable execution of quantum solvers, making for an adequate evaluation of their exploration quality mandatory execution on quantum simulators. Since simulators' complexity, and, consequently, execution time,  grows exponentially with the number of involved qubits, a fair comparison with QA and SA is not permitted. For this reason, this study focuses only on the quality of the achieved solution, with execution time considerations left for future exploration when quantum computers' reliability improves within the quantum circuit model.

Given the stochastic nature of solvers, evaluating their outcomes requires \textit{multiple executions} to assess solution quality comprehensively. In this work, we considered one hundred runs of each algorithm for every solver. We believed this number sufficient to gather meaningful statistical insights while avoiding excessive computational demands that could make the tests impractical without sophisticated high-performance computing platforms. 

In order to consolidate various metrics related to the distribution of the overall results---i.e., the minimum value, the average and the variance of the obtained results, the percentage of optimal/suboptimal solutions, etc.---into a single figure of merit, we propose a scoring system composed of weighted indicators to minimize:
\vspace{-4pt}
\begin{equation}
    F_{\textrm{s}} = -\alpha p_{\textrm{s}} + \beta (E_{\textrm{opt}} - E_{\textrm{ref}}) + \gamma (E_{\textrm{avg}} - E_{\textrm{ref}}) + \delta E_{\textrm{var}} - \eta p_{\textrm{v}} \, , \vspace{-4pt}
\end{equation} 
where $p_{\textrm{s}}$ represents the percentage of outcomes equal to the optimum, $E_{\textrm{opt}}$ denotes the best-achieved value, $E_{\textrm{ref}}$ stands for the reference optimal value for the problem, $E_{\textrm{avg}}$ signifies the average value obtained, $E_{\textrm{var}}$ indicates the variance of the obtained values, and $p_{\textrm{v}}$ is the percentage of solutions satisfying the constraints. $\alpha$, $\beta$, $\gamma$, $\delta$, and $\eta$ represent tunable weights. This study considers all metrics equally important, with each weight assigned to the same value.

Once the figure of merit for solver performance evaluation is defined, a policy for handling cases of \textit{parity scores} among the solvers must be established. One approach is to develop a multi-label, multi-class classifier, enabling the assignment of multiple labels to dataset items. Alternatively, a priority map among the solvers can be defined based on their characteristics. For simplicity in the current implementation, the latter approach is adopted. The \textit{preference criterium} is based on the \textit{number of qubits} involved in the solver and a preference for quantum solutions over the classical one. Specifically, the priority order is as follows:
\begin{enumerate}
\item QAOA, involving a number of qubits equal to the number of QUBO variables;
\item VQE, also involving a number of qubits equal to the number of QUBO variables;
\item GAS, involving a number of qubits equal to the number of QUBO variables plus the number of bits for the function operative range representation;
\item QA, involving multiple qubits for each binary variable due to minor embedding;
\item SA, which is the unique classical alternative considered. 
\end{enumerate}
In establishing the priority order, we have considered QAOA before VQE, even if they required the same number of qubits, based on empirical observations. Indeed, we have noticed that, on average, QAOA reaches convergence faster than VQE.


\subsubsection{Identification of the Problems Features: }
Afterwards, it is necessary to define QUBO information with minimal computational effort for their inclusion as \textit{features} in the prediction model. Considering the characteristics of the QUBO solver, we proposed to employ the following features: 
\begin{itemize}
    \item number of variables in the QUBO problem;
    \item number of non-zeros first-order QUBO terms ($a_i$); 
    \item number of non-zeros second-order QUBO coefficients ($b_{ij}$);
   \item average of non-zeros first-order QUBO terms ($a_i$);
   \item variance of non-zeros first-order coefficients ($a_i$);
   \item average of non-zeros second-order QUBO coefficients ($b_{ij}$);
   \item variance of non-zeros second-order QUBO terms ($b_{ij}$);
   \item average of all coefficients ($a_i$ and $b_{ij}$);
   \item variance of all coefficients ($a_i$ and $b_{ij}$).
\end{itemize}
One of the main challenges in this step lies in identifying problem characteristics that may affect solver performance and estimating them with limited computational resources. Specifically, we focused on the count of non-zero coefficients of the second order, as they influence the number of two-qubit relationships that need to be expressed on quantum computers. The total number of variables, as well as the counts of first and second-order elements, collectively determine the QUBO density, which affects the shape of the associated cost function. Moreover, the average and variance of the problem coefficients impact the fluctuations' width in the QUBO function, thus also influencing the effectiveness of various exploration strategies, as suggested in \cite{chancellor2017modernizing}.

All this information can be easily extrapolated from the final QUBO formulation with computational complexity growing at most quadratically with the number of problem variables.
\vspace{8pt}
\subsubsection{Creation of the Dataset:}
After the definition of evaluation metrics, priority map and features, it is necessary to generate a proper dataset for training and evaluating the classification models. More than \textit{500 different QUBO problems} have been considered. In particular, we consider: 
\begin{itemize}
 \item the well-known benchmark of conventional optimization (those without non-linear elements) reported in~\cite{surjanovic_optimization} and written as QUBO;
\item problems generated of different sizes, densities and range of coefficients from the QUBO formulations  reported in~\cite{glover2018tutorial};
\item problems generated of different sizes, densities and range of coefficients from the QUBO formulations reported in~\cite{lucas2014ising};
\item portfolio optimization problem, written as QUBO;
\item linear regression problem on the Iris dataset written as QUBO.
\end{itemize}

These benchmarks have been carefully selected to encompass a wide range of optimization problems that users may encounter, providing a comprehensive overview of potential scenarios of interest.

\subsubsection{Preprocessing Techniques:}\label{sec:Preprocessing}
The resulting dataset exhibits class \textit{imbalance}, meaning that the proportion of samples belonging to each class changes. In particular, QAOA emerges as the best solver for little less than half of the entire dataset,  followed by QA and VQE, each representing approximately 20\%,  while GAS and SA are optimal for around 10\% of the analyzed problems. 
Unfortunately, an unbalanced classification task is more complex than a balanced one~\cite{brownlee2020imbalanced}.
\begin{figure}[h]
    \centering\vspace{-10pt}
    \includegraphics[width=0.85\linewidth]{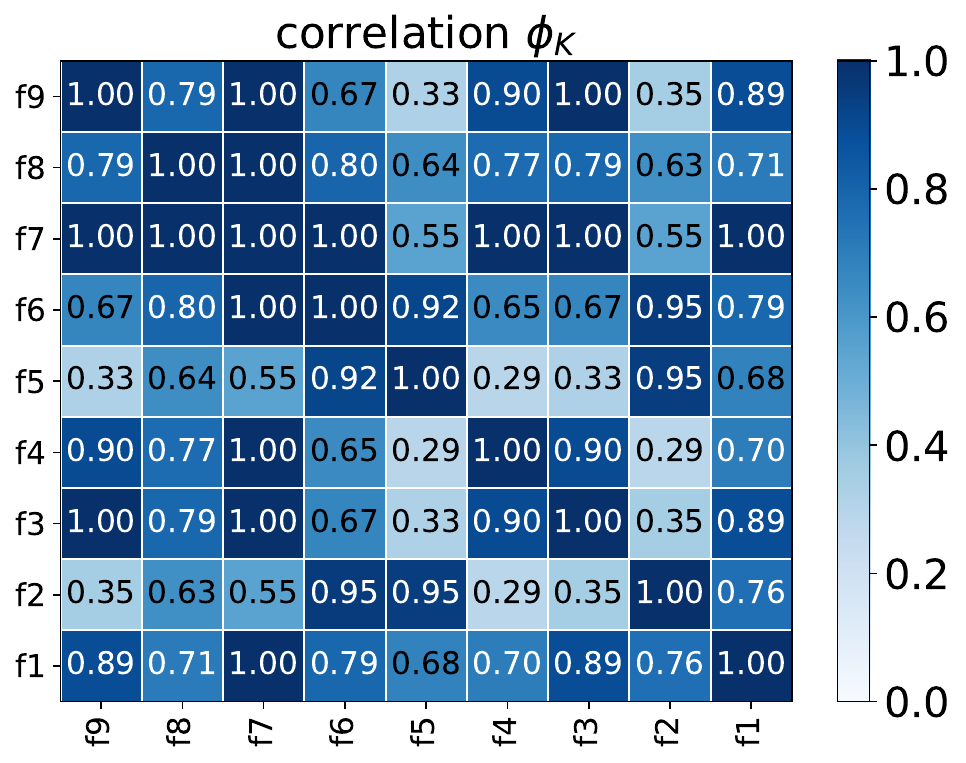}\vspace{-8pt}
    \caption{$\phi_K$ correlation matrix defined in~\cite{baak2020new}. It expresses also non-linear relations among the features. \vspace{-8pt}}
    \label{fig:Correlation}
\end{figure}

Some \textit{pre-processing} techniques can be applied to improve model stability and \textit{reduce the dataset's redundancy and dimensionality}, mitigating the task complexity. First of all, features have been scaled to a common range to avoid model domination by the larger scale features and to remove model bias. Moreover, in this work, we evaluate the potential of the \textit{Principal Component Analysis} (PCA,~\cite{jolliffe2016principal}) and \textit{Linear Discriminant Analysis} (LDA,~\cite{tharwat2017linear}) techniques.
 PCA, an unsupervised learning method, identifies the principal components, representing orthogonal directions of maximum variance, to capture the dataset's greatest variability. On the other hand, LDA, a supervised learning technique, computes the optimal data projection onto a lower-dimensional subspace, maximizing class separability. Both techniques aim to extract features while retaining crucial information in the data.

In this study, we explore datasets that have been dimensionally reduced using PCA with 2, 3, 4 (retaining 99\% of data variability), and 9 components, alongside LDA with 2, 3, and 4 elements. Examination of the $\phi_K$ correlation matrix presented in Figure~\ref{fig:Correlation} suggests that PCA and LDA  could have a positive impact on model performance since the dataset presents redundancy and relations among features.

\subsubsection{Supervised Learning Models:}\label{sec:models}
For the predictor development, a five-fold cross-validation technique has been exploited, considering the following supervised machine-learning classifiers of the scikit-learn library~\cite{scikit-learn}, properly tuning their degrees of freedom: 
\vspace{-3pt}
\begin{itemize}
    \item \textit{Ada Boost}~\cite{ferreira2012boosting};
    \item \textit{Decision Tree}~\cite{safavian1991survey};
    \item \textit{Gradient Boosting}~\cite{he2019gradient};
    \item \textit{k-nearest neighbors} (KNN,~\cite{bhatia2010survey});
    \item \textit{Logistic Regression}~\cite{das2017survey};
    \item \textit{Naive Bayes}~\cite{das2017survey};
    \item \textit{Neural Network}~\cite{smys2020survey};
    \item Random Forest~\cite{zakariah2014classification};
    \item \textit{Support Vector Machine} (SVM,~\cite{cervantes2020comprehensive});
    \item \textit{eXtreme Gradient Boosting} (XGBoost~\cite{chen2016xgboost}).
\end{itemize}
\vspace{-3pt}
To address the dataset's imbalance, the creation of the folds has exploited a \textit{stratification technique} for maintaining the proportion among elements belonging to each class. 

The \mbox{k cross-validation} approach is crucial for optimizing model performance and ensuring robustness and reliability in estimating the quality of solver prediction with unseen problems. Additionally, it enhances efficiency in comparing the generalization capabilities of different models.
\vspace{-5pt}
\subsection{Solvers Parameter Setting} \label{sec:setting}
\vspace{-3pt}
In this section, simple strategies for effectively configuring and scaling the parameters of quantum solvers are proposed. These options are those considered for the predictive model dataset generation and, consequently, for each new problem for configuring the selected solver. Specifically, for the dataset creation, the \textit{Advantage\_system4.1} D-Wave device has been chosen for QA, while the Qiskit implementation of QAOA, VQE, and GAS solvers, launched on the QASM simulator, has been employed for their respective configurations. Finally, the D-Wave implementation of the SA solver has been employed.
\subsubsection{Quantum Annealer: }
Defining the \textit{annealing time} scaling with the QUBO problem size is crucial for QA. This relation can be deduced from state-of-the-art plots representing the \textit{\mbox{time-to-solution}} (TTS) metric relative to the problem size. In~\cite{mandra2016strengths}, it is proved that the annealing time necessary is expected to grow as
\vspace{-10pt}
\begin{equation}
    T_{\textrm{TTS}} = 10^{b\sqrt{N}} \, , \vspace{-5pt}
    \label{eq:TTS}
\end{equation}
where $N$ is the number of variables, while $b$ is a coefficient. From plots in the literature~\cite{mandra2016strengths} and our experience, $b$ equals 0.7 is a reasonable value.
\subsubsection{Quantum Approximate Optimization Algorithm: }
The parameter most critical to scale with the problem size in QAOA is the number of \textit{repetitions}, as discussed in Section~\ref{sec:Quantum solver}. Also in this case, the scaling function for this parameter has been derived from literature plots, particularly those found in~\cite{lee2021parameters,niu2019optimizing}. Our analysis reveals that the number of repetitions required for achieving satisfactory results increases proportionally to $\sqrt{N}$, where $N$ represents the number of problem variables. The relation adopted for dataset creation is given by:\vspace{-7pt}
\begin{equation*}
\textrm{reps} = \ceil{2\sqrt{N}} \, . \vspace{-5pt}
\end{equation*} 
Moreover, we have considered Hadamard state initialization and $-X$ mixed state, as in Figure~\ref{fig:QAOAExample}. For classical optimization, the Cobyla optimizer, providing on average good performance, has been selected without imposing restrictions on the number of classical iterations, allowing the algorithm to automatically stop when the convergence is achieved.

\subsubsection{Variational Quantum Eigensolver: }
Unlike the other solver, VQE has no critical parameters to be set scaling with the problem size. For dataset creation, an ansatz composed of a $Ry$ gate layer and an entangling layer, like that shown in Figure~\ref{fig:VQEExample}, has been chosen. Similarly to QAOA, for classical optimization, Cobyla, providing on average good performance, has been exploited without a restriction on the number of classical iterations, allowing the algorithm to terminate upon convergence.
\subsubsection{Grover Adaptive Search: }
The GAS parameter setting is very complex, involving the determination of the algorithm's stop \textit{threshold}, which scales with the problem size, and the selection of the \textit{number of qubits for representing cost function values}, dependent on the specific problem.

Unfortunately,  analyses for threshold scaling are lacking in the literature. Therefore, we have empirically discovered that a linear growth with the problem size provides reasonable results quality, and it is the same scaling adopted in~\cite{giuffrida2022engineering} with the Qiskit framework. In particular, for dataset creation, the following scaling function has been considered:  \vspace{-7pt}
\begin{equation}
\textrm{th} = 2N \, ,  \vspace{-5pt}
\end{equation}
where $N$ is the number of problem variables. 

For what concerns cost function values encoding, we propose to leverage the estimation of the \textit{functions bound} based on the exploitation of \textit{posiform} and \textit{negaform} proposed in~\cite{boros2002pseudo, boros2006preprocessing} after the application of a preprocessing step. Considering that the image of the cost function is represented in fixed point, for saving qubits, we choose first to round the problem coefficients to a user-selected precision, then \textit{normalize} all QUBO matrix elements such that the smallest one (in terms of absolute value) is equal to 1. In this way, the waste in terms of the number of value qubits should be minimized. 

\subsubsection{Simulated Annealing: }
Similarly to QA, for SA, the crucial parameter to scale with the problem size is the \textit{number of iterations} (or sweeps), which is strongly related to annealing time. Also in this case, TTS can be exploited, and the same type of scaling of QA (Equation~\ref{eq:TTS}) is expected from analysis in~\cite{mandra2016strengths}.
Empirically, we found that $b$ equals 0.5 is a reasonable value. 
{\renewcommand{\arraystretch}{1.6}
\setlength{\tabcolsep}{3pt}
\begin{table}[t]
\caption{Performance comparison across various classifiers and feature reduction methods, highlighting accuracy (Acc), the percentage of top two predictions, and the average error in the success probability ($p_s$ err). The best results are highlighted in \textbf{bold} and \textcolor{teal}{green}. \vspace{-10pt} }\label{tab:Prediction}\begin{center}
\resizebox{0.5\textwidth}{!}{
\begin{tabular}{?c|c?c?c|c|c|c?c|c|c?}
\noalign{\hrule height 1.5pt}
\multicolumn{2}{?c?}{\multirow{2}{*}{\textbf{Models}}} & \multirow{2}{*}{\textbf{No preproc}}  & \multicolumn{4}{c?}{\textbf{PCA}}  & \multicolumn{3}{c?}{\textbf{LDA}}\\  \cline{4-10}
 \multicolumn{2}{?c?}{} &  & \textbf{2} & \textbf{3} & \textbf{4} & \textbf{9} & \textbf{2} & \textbf{3} & \textbf{4} \\
\noalign{\hrule height 1.5pt}
\multirow{3}{*}{\textbf{Ada Boost}} & \textbf{Accuracy [\%]}&64.48&55.79&59.78&62.68&62.85&64.30&64.30&65.21\\
  \cline{2-10}  
 & \textbf{Top two [\%]}&86.23&81.70&82.07&85.14&85.33&84.96&84.78&84.96\\
  \cline{2-10}  
 & \textbf{$p_s$ err [\%]}&4.26&6.90&6.87&7.22&7.42&7.31&7.57&6.91\\
\noalign{\hrule height 1.5pt} 
\multirow{3}{*}{\textbf{Decision Tree}} & \textbf{Accuracy [\%]}&68.65&64.85&65.75&63.57&67.02&65.75&65.75&65.93\\
  \cline{2-10}  
 & \textbf{Top two [\%]}&87.50&84.60&85.87&84.78&85.14&85.33&85.33&85.69\\
  \cline{2-10}  
 & \textbf{$p_s$ err [\%]}&3.22&4.79&4.45&4.45&4.91&4.80&4.31&4.42\\
\noalign{\hrule height 1.5pt} 
\multirow{3}{*}{\textbf{Gradient Boosting}} & \textbf{Accuracy [\%]}&72.63&67.02&67.75&69.20&69.37&66.66&66.29&66.65\\
  \cline{2-10}  
 & \textbf{Top two [\%]}&89.86&86.41&87.50&86.96&86.59&85.87&86.05&86.78\\
  \cline{2-10}  
 & \textbf{$p_s$ err [\%]}&2.40&4.37&4.67&4.15&4.37&4.80&4.64&3.89\\
\noalign{\hrule height 1.5pt} 
\multirow{3}{*}{\textbf{KNN}} & \textbf{Accuracy [\%]}&57.79&56.52&56.88&57.42&57.42&59.24&59.24&59.96\\
  \cline{2-10}  
 & \textbf{Top two [\%]}&81.52&78.08&78.44&79.89&79.89&82.79&82.25&82.79\\
  \cline{2-10}  
 & \textbf{$p_s$ err [\%]}&7.37&7.07&7.18&7.33&7.33&6.86&6.97&6.92\\
\noalign{\hrule height 1.5pt} 
\multirow{3}{*}{\textbf{Logistic Regression}} & \textbf{Accuracy [\%]}&71.01&32.06&42.57&50.18&50.17&53.62&55.06&55.43\\
  \cline{2-10}  
 & \textbf{Top two [\%]}&88.59&55.25&70.29&78.26&77.72&78.26&80.80&81.70\\
  \cline{2-10}  
 & \textbf{$p_s$ err [\%]}&3.70&13.32&9.62&7.64&8.17&9.96&6.60&6.31\\
\noalign{\hrule height 1.5pt} 
\multirow{3}{*}{\textbf{Naive Bayes}} & \textbf{Accuracy [\%]}&53.09&56.52&59.41&61.58&60.49&60.68&60.32&60.68\\
  \cline{2-10}  
 & \textbf{Top two [\%]}&77.36&78.08&81.52&83.88&84.06&82.43&82.43&83.51\\
  \cline{2-10}  
 & \textbf{$p_s$ err [\%]}&7.49&7.07&7.02&6.70&7.37&8.66&8.05&7.89\\
\noalign{\hrule height 1.5pt} 
\multirow{3}{*}{\textbf{Neural Network}} & \textbf{Accuracy [\%]}&57.78&63.76&63.94&66.12&64.49&63.04&64.48&63.58\\
  \cline{2-10}  
 & \textbf{Top two [\%]}&81.16&85.51&87.14&85.87&86.23&87.50&87.32&86.59\\
  \cline{2-10}  
 & \textbf{$p_s$ err [\%]}&6.63&5.69&6.52&6.73&7.13&6.50&6.90&6.81\\
\noalign{\hrule height 1.5pt} 
\multirow{3}{*}{\textbf{Random Forest}} & \textbf{Accuracy [\%]}&\textcolor{teal}{\textbf{73.18}}&66.30&68.46&69.92&68.83&69.56&68.29&68.65\\
  \cline{2-10}  
 & \textbf{Top two [\%]}&\textcolor{teal}{\textbf{91.12}}&86.05&87.68&88.95&87.50&88.59&87.32&87.32\\
  \cline{2-10}  
 & \textbf{$p_s$ err [\%]}&\textcolor{teal}{\textbf{2.16}}&4.06&3.80&3.17&3.87&3.70&3.80&3.94\\
\noalign{\hrule height 1.5pt} 
\multirow{3}{*}{\textbf{SVM}} & \textbf{Accuracy [\%]}&59.06&56.52&61.04&62.68&62.68&63.04&64.48&63.94\\
  \cline{2-10}  
 & \textbf{Top two [\%]}&83.70&78.08&84.24&85.69&85.69&86.96&87.14&86.59\\
  \cline{2-10}  
 & \textbf{$p_s$ err [\%]}&7.17&7.07&6.52&6.62&6.62&6.55&6.46&6.43\\
\noalign{\hrule height 1.5pt} 
\multirow{3}{*}{\textbf{XGBoost}} & \textbf{Accuracy [\%]}&69.56&66.30&67.02&68.46&66.84&66.12&67.19&67.02\\
  \cline{2-10}  
 & \textbf{Top two [\%]}&87.50&85.69&86.23&87.50&84.96&87.14&87.50&87.50\\
  \cline{2-10}  
 & \textbf{$p_s$ err [\%]}&3.01&4.12&4.57&3.66&4.61&4.47&4.24&4.05\\
\noalign{\hrule height 1.5pt} 
\end{tabular}}
\end{center} \vspace{-20pt}
\end{table}
\vspace{-10pt}
\section{Results} \label{sec:Results}
The supervised learning models, presented in Section \ref{sec:models}, have been trained and evaluated with and without applying the features reduction technique discussed in Section \ref{sec:Preprocessing} exploiting a five-fold cross-validation procedure on a dataset including more than 500 different QUBO problems. 
The classifier and feature reduction techniques quality are compared in Table \ref{tab:Prediction} in terms of:
\begin{itemize}
    \item \textit{Accuracy}, which is the percentage of correct predictions found with the k cross-validation approach.
    \item \textit{Top 2}, which is the relative frequency of predicting one of the top two solvers.
    \item \textit{Average $p_s$ error}, which is the average distance between the probability of achieving the optimal solution of the best solver and of the predicted solver. 
\end{itemize}
We choose to present the average error in success probability rather than those in the overall score because it offers a more explicit interpretation with respect to an abstract score. Furthermore, its range of possible values is problem-independent, unlike the score---dependent on function bounds---, thus simplifying the process of averaging the results.

The effectiveness of the features reduction technique depends on the type of model. For example, considering the model SVM, Naive Bayes, or Neural Network, they are particularly useful, while in  Random Forest, they do not provide benefit.

Observing the table, it is clear that Random Forest outperforms other supervised learning models in predicting the expected optimal solver. It achieves the best performance with an accuracy of 73.18\% and about 90\% rate of predictions providing a solver in the top two. Additionally, the average loss in terms of the probability of obtaining the optimal solution is almost negligible, at just 2.16\%. These results have been obtained with the following hyperparameter setting:
\begin{itemize}
    \item number of Decision Trees equal to 100;
    \item maximal depth equal to 50;
    \item minimal samples per leaf equal to 1;
\item minimal samples per split equal to 2.
\end{itemize}

\begin{figure}[h]
    \centering
    \vspace{-10pt}\includegraphics[width=0.85\linewidth]{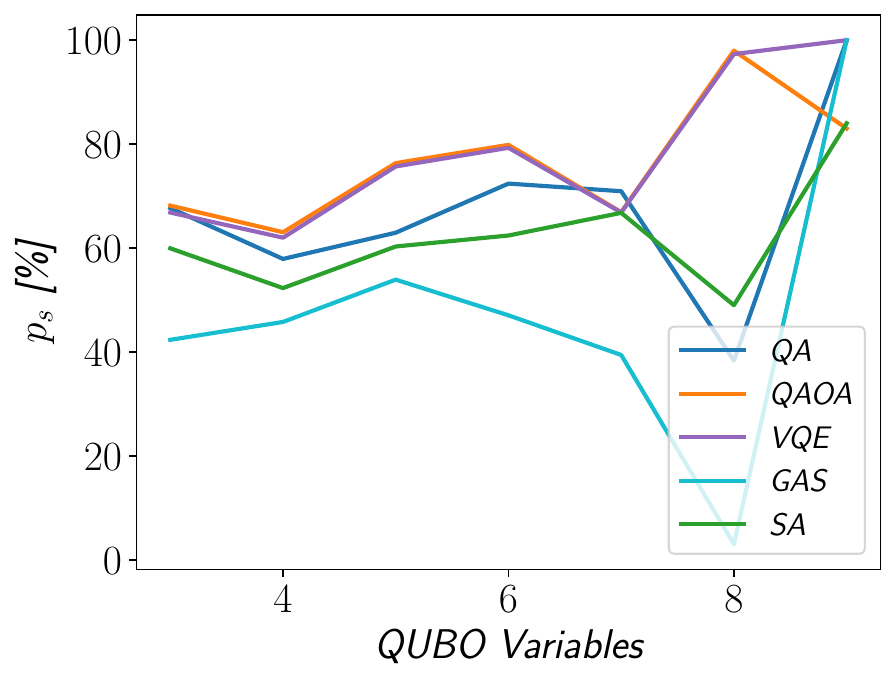}\vspace{-8pt}
    \caption{ 
The average probability of achieving the expected optimal value ($p_s$) across the dataset's problems is plotted against the number of QUBO variables. \vspace{-8pt}}
    \label{fig:Setting}
\end{figure}
Finally, Figure \ref{fig:Setting} shows the average probability of achieving the optimal solution of the problems considered in the dataset creation across their sizes. These problems, as previously mentioned, exhibit considerable variations in density,  dimension, and variance of the involved coefficients, etc. It is possible to notice the consistency in solution quality across different problem sizes, proving the reasonableness of the proposed parmeters scaling approach.

\vspace{-3pt}
\section{Discussion}\label{sec:Discussion}
\vspace{-3pt}
The conducted exploration demonstrates the feasibility of approaching solver selection as a classification task. However, a significant challenge lies in the necessity of a large dataset, which is computationally and economically expensive to obtain, to ensure a reliable prediction. While the dataset dimensions in this manuscript suffice to establish proof of concept, providing encouraging results, more training data are needed to achieve completely satisfactory prediction results. Expanding the dataset poses challenges related to the selection of diverse problems to provide a comprehensive overview of potential scenarios. Furthermore, the maximum size of the problems that can be addressed is constrained by the limits of current quantum computer simulators, especially for GAS execution, requiring the use of complex High-Performance Computing systems to be able to consider the most interesting problems.

Moreover, while the chosen solvers' settings for dataset creation have demonstrated their reasonableness, relying exclusively on empirical deductions is not the optimal approach. Consequently, there is potential for significant enhancements in the implementation of predictors.

\vspace{-3pt}
\section{Conclusions}\label{sec:conclusions}
\vspace{-3pt}
This work proposes to address the \textit{solver selection} challenge with a \textit{supervised machine learning approach}, treating it as a classification task. Moreover, the article suggests strategies for \textit{adjusting solver parameters} based on problem size and characteristics. To this end, we first reviewed quantum optimization, focusing on solvers and flow required for solving an optimization problem with quantum computers. Then, we delve into the motivation driving this exploration and the objectives of the research. Afterwards, the methodology followed for developing the solver predictor is outlined step by step, explaining and motivating design choices, and the solvers' setting, based on state-of-the-art experience and considered for dataset creation, has been introduced. The effectiveness of the supervised learning approach is validated through experimentation with a \textit{dataset comprising 500 diverse QUBO problems}.

The pre-trained classifier is integrated into the MQT Quantum Auto Optimizer (MQT QAO) framework,  publicly available on GitHub (\url{https://github.com/cda-tum/mqt-qao}) as part of the Munich Quantum Toolkit (MQT).

Even though the obtained results are promising, opportunities for enhancement and expansion remain. First, the proposed parameters' setting can be improved by involving machine learning models. Moreover, the solver selection predictor can be improved both enlarging the considered training dataset and considering more quantum and classical solvers and their variances. In addition, a \mbox{multi-label} approach for managing the parity cases, instead of the priority mechanism, can be considered, and reinforcement learning can be evaluated as an alternative to supervised learning. 

In conclusion, this research proves the potential of machine learning in quantum solver selection,  offering valuable tools for non-experts in quantum computing. We hope that this exploration will be the starting point towards the development of automated instruments for managing quantum solvers, thereby enabling the creation of quantum solutions for \mbox{real-world} problems by a broader spectrum of users.

\vspace{-3pt}
\section*{acknowledgements}
\vspace{-3pt}
N.Q. and R.W. acknowledge funding from the European Research Council (ERC) under the European Union’s Horizon 2020 research and innovation program (grant agreement No. 101001318), the Munich Quantum Valley, which is supported by the Bavarian state government with funds from the Hightech Agenda Bayern Plus, and the BMWK on the basis of a decision by the German Bundestag through project QuaST, as well as the BMK, BMDW, the State of Upper Austria in the frame of the COMET program, and the QuantumReady project within Quantum Austria (managed by the FFG).\\
D.V.  would like to thank HPC@POLITO --- a project of Academic Computing within the Department of Control and Computer Engineering at the Politecnico di Torino (\url{http://hpc.polito.it}) --- for providing computational resources.

\bibliographystyle{ieeetr}
\bibliography{sn-bibliography}

\end{document}